\documentclass[acmtog]{acmart}
\newcommand{\validation}[1]{\textcolor{magenta}{\footnotesize #1}} 
\usepackage{nicefrac}
\usepackage{multirow}
\usepackage{graphicx}
\usepackage{subfigure}
\usepackage{nicefrac}
\usepackage{placeins}
\usepackage{letltxmacro}
\usepackage[nolist]{acronym}

\begin{acronym}
\acro{VRR}{variable refresh rate}
\acro{CFF}{Critical Flicker Frequency}
\acro{TLM}{temporal light modulation}
\acro{TCSF}{Temporal Contrast Sensitivity Function}
\acro{CSF}{Contrast Sensitivity Function}
\acro{fMRI}{functional magnetic resonance imaging}
\acro{ICDM}{International Committee Display Metrology}
\acro{IDMS}{Information Display Measurements Standard}
\acro{GPU}{Graphics Processing Unit}
\acro{V-blank}{Vertical Blanking Interval}
\acro{RMSE}{root-mean-square-error}
\end{acronym}

\AtBeginDocument{%
  \providecommand\BibTeX{{%
    \normalfont B\kern-0.5em{\scshape i\kern-0.25em b}\kern-0.8em\TeX}}}

\copyrightyear{2024}
\acmYear{2024}
\setcopyright{rightsretained}
\acmConference[SA Conference Papers '24]{SIGGRAPH Asia 2024 Conference
Papers}{December 3--6, 2024}{Tokyo, Japan}
\acmBooktitle{SIGGRAPH Asia 2024 Conference Papers (SA Conference Papers
'24), December 3--6, 2024, Tokyo, Japan}\acmDOI{10.1145/3680528.3687586}
\acmISBN{979-8-4007-1131-2/24/12}
  
\acmSubmissionID{297}

\begin{document}
\newcommand{\figref}[1]{Figure~\ref{fig:#1}}
\newcommand{\secref}[1]{Section~\ref{sec:#1}}
\newcommand{\algoref}[1]{Algorithm~\ref{algo:#1}}
\newcommand{\chapterref}[1]{Chapter~\ref{chapter:#1}}
\newcommand{\appref}[1]{Appendix~\ref{app:#1}}
\newcommand{\tableref}[1]{Table~\ref{tab:#1}}
\newcommand{\etal}{et al.\xspace}
\newcommand{\degpers}{\,\nicefrac{deg}{s}\xspace}
\newcommand{\degperssq}{\,\nicefrac{deg}{s\textsuperscript{2}}\xspace}
\newcommand{\degree}{$^{\circ}$\xspace}
\newcommand{\Hz}{\,Hz\xspace}
\newcommand{\csdm}{\,cd/m$^2$}
\newcommand{\fourier}{\mathfrak{F}}
\newcommand{\infourier}{^{\mathfrak{F}}}

\LetLtxMacro{\originaleqref}{\eqref}
\renewcommand{\eqref}[1]{Eq.~\originaleqref{eq:#1}}

\newcommand{\todo}[1]{\textcolor{red}{\textbf{todo: #1}}}
\newcommand{\RM}[1]{\textcolor{brown}{\textnormal{(Rafal) #1}}}
\newcommand{\AB}[1]{\textcolor{purple}{\textnormal{(Ali) #1}}}
\newcommand{\MA}[1]{\textcolor{blue}{\textnormal{(Maliha) #1}}}
\newcommand{\MAtext}[1]{\textcolor{blue}{\textnormal{#1}}}

\newcommand{\edit}[1]{\textcolor{black}{{#1}}}

\newcommand{\ourmethod}{elaTCSF}

\newcommand{\code}[1]{\texttt{#1}}

\newcommand{\ind}[1]{\text{#1}}


\title{elaTCSF: A Temporal Contrast Sensitivity Function for Flicker Detection and Modeling Variable Refresh Rate Flicker}

\author[]{Yancheng Cai}
\email{yc613@cam.ac.uk}
\affiliation{
  \institution{University of Cambridge}
  \streetaddress{William Gates Building, 15 JJ Thomson Avenue}
  \city{Cambridge}
  \postcode{CB3 0FD}
  \country{United Kingdom}
}

\author[]{Ali Bozorgian}
\email{ali.bozorgian@ntnu.no}
\affiliation{
  \institution{Norwegian University of Science and Technology}
  \city{Gjøvik}
  \country{Norway}
}

\author[]{Maliha Ashraf}
\email{ma905@cam.ac.uk}
\affiliation{
  \institution{University of Cambridge}
  \streetaddress{William Gates Building, 15 JJ Thomson Avenue}
  \city{Cambridge}
  \postcode{CB3 0FD}
  \country{United Kingdom}
}

\author[]{Robert Wanat}
\email{robwanat@gmail.com}
\affiliation{
  \institution{LG Electronics North America}
  \city{Santa Clara}
  \country{United States of America}
}

\author[]{Rafał K. Mantiuk}
\email{rafal.mantiuk@cl.cam.ac.uk}
\affiliation{
  \institution{University of Cambridge}
  \streetaddress{William Gates Building, 15 JJ Thomson Avenue}
  \city{Cambridge}
  \postcode{CB3 0FD}
  \country{United Kingdom}
}


\begin{abstract}
The perception of flicker has been a prominent concern in illumination and electronic display fields for over a century. Traditional approaches often rely on Critical Flicker Frequency (CFF), primarily suited for high-contrast (full-on, full-off) flicker. To tackle varying contrast flicker, the International Committee for Display Metrology (ICDM) introduced a Temporal Contrast Sensitivity Function TCSF$_{IDMS}$ within the Information Display Measurements Standard (IDMS). Nevertheless, this standard overlooks crucial parameters: luminance, eccentricity, and area. Existing models incorporating these parameters are inadequate for flicker detection, especially at low spatial frequencies. To address these limitations, we extend the TCSF$_{IDMS}$ and combine it with a new spatial probability summation model to incorporate the effects of luminance, eccentricity, and area (elaTCSF). We train the elaTCSF on various flicker detection datasets and establish the first variable refresh rate flicker detection dataset for further verification. Additionally, we contribute to resolving a longstanding debate on whether the flicker is more visible in peripheral vision. We demonstrate how elaTCSF can be used to predict flicker due to low-persistence in VR headsets,  identify flicker-free VRR operational ranges, and determine flicker sensitivity in lighting design.

\end{abstract}



\begin{teaserfigure}
  \includegraphics[height=4.52cm]{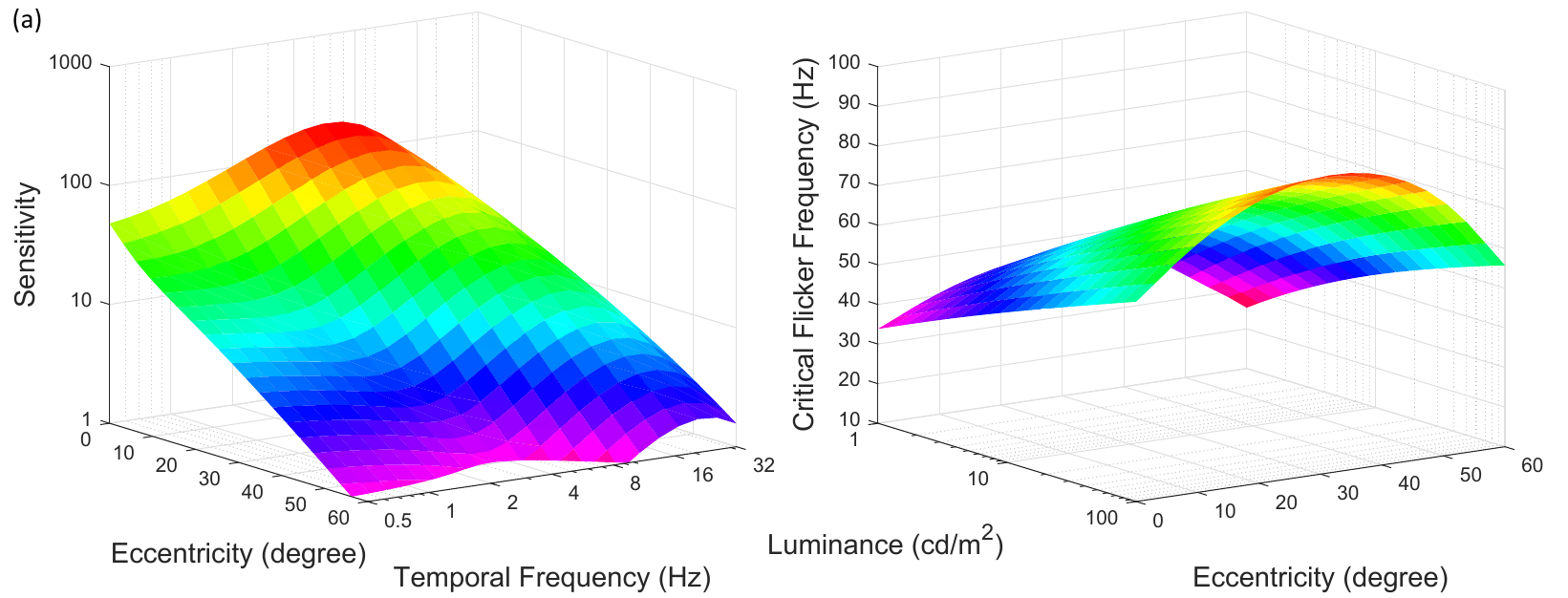}
  \includegraphics[height=4.52cm]{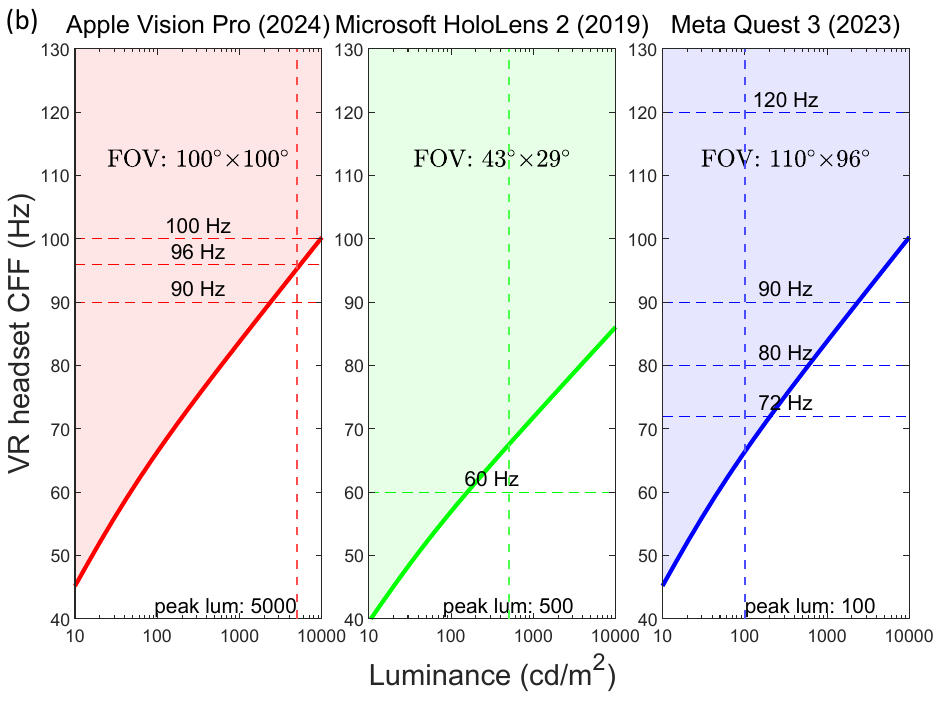}
  \caption{(a) 3D Visualization of elaTCSF. elaTSCF predicts sensitivity and critical flicker frequency along three dimensions: Eccentricity, Luminance, and Area. (b) elaTCSF predicts flicker visibility in short-persistence displays, such as VR/AR headsets. The dashed lines indicate the current display capabilities (peak luminance and official refresh rate). Above the CFF threshold (colored area), flicker is unlikely to be visible.
  }
  \label{fig:teaser}
\end{teaserfigure}


\maketitle

\section{Introduction}
\label{sec:introduction}

Flicker, known as \ac{TLM}, has been a significant concern in illumination engineering for over a century. It has detrimental effects on human physiology, ranging from irritation to neurological disturbances~\cite{miller2023flicker}. Flicker has been observed in different technologies such as fluorescent lamps~\cite{eastman1952stroboscopic} in the 1940s, cathode-ray tubes (CRTs)~\cite{bauer1983frame} in the 1980s, and LED~\cite{lehman2011proposing}.


The \ac{CFF} is the best-known flicker detection standard that has been in use for over a century~\cite{porter1902contributions}, defined as the frequency at which a flickering light becomes indistinguishable from a steady, non-flickering light. However, it assumes a high-contrast flicker (light on and off), which is not always applicable. For example, flicker found in \ac{VRR} displays is caused by slight luminance differences at varying refresh rates, resulting in a low-contrast flicker. To address flicker of various contrast, Watson proposed a flicker detection metric~\cite{watson201164}, which relies on the \ac{TCSF} equation from \cite{watson1986temporal}. This metric (TCSF$_{IDMS}$\footnote{TCSF describes human visual sensitivity as a function of temporal frequency, exhibiting diverse forms. TCSF$_{IDMS}$ specifically denotes the standard proposed by the International Committee for Display Metrology (https://www.sid.org/Standards/ICDM).}) was later incorporated into the information display measurements standard (IDMS) v1.2 by the \ac{ICDM} and was released in 2023. However, as acknowledged by  \citeN{watson201164}, other factors such as luminance, area, and eccentricity also affect the shape of \ac{TCSF}, yet these parameters have not been integrated into the current standard, limiting its applications. Existing \acp{CSF}, such as Barten's~\citeNN{barten1999contrast}, stelaCSF~\cite{mantiuk2022stelacsf} and castleCSF~\cite{ashraf2024castlecsf} predict sensitivity as a function of mentioned factors, but (a) do not offer sufficient accuracy at high temporal frequencies for modeling flicker; (b) cannot explain flicker visibility for large sources that span a significant portion of the visual field. This paper aims to extend the TCSF$_{IDMS}$ to incorporate the effects of eccentricity, luminance and area (hence elaTCSF), updating the standard for flicker detection.

A 120-year-long debate persists in flicker detection: Is temporal sensitivity to flicker higher in the peripheral visual field (parafovea) than in the central (fovea)? The organization of the visual field is topographically represented in the striate cortex, with central regions receiving a disproportionately larger representation than peripheral areas~\cite{daniel1961representation}. However, regarding temporal sensitivity, experimental measurements are conflicting and puzzling: some researchers observe a rise and fall in \ac{CFF} with increasing eccentricity, peaking in the parafovea~\cite{porter1902contributions, phillips1933perception, hylkema1942examination, hartmann1979peripheral, rovamo1984critical, tyler1987analysis, krajancich2021perceptual}, while others contend that temporal sensitivity peaks at fovea and decreases steadily with eccentricity~\cite{ross1936fusion, chapiro2023critical}. Presently, a consensus remains elusive, although the data from \citeN{hartmann1979peripheral} shows that larger stimulus areas, higher luminances, and higher contrasts are more likely to elicit peaks in the parafovea. We rely on those findings, including modern \ac{fMRI} measurements of cortex activity \cite{horiguchi2009two, himmelberg2019eccentricity}, and propose \edit{a model that can explain these seemingly conflicting observations}. 

Larger signal areas generally activate a broader range of retinal cone and rod cells, enhancing human sensitivity. Some CSF models like castleCSF~\cite{ashraf2024castlecsf} have incorporated area (size) as a crucial parameter. However, they treat area and eccentricity as independent parameters. In reality, for stimuli with large areas, a range of eccentricities is involved, particularly evident in our VRR flicker dataset, which includes full-screen flicker. Relying solely on one eccentricity value is evidently insufficient. In this context, we propose a spatial probability summation model that can work with stimuli spanning any portion of the visual field. 

In summary, our contributions are as follows:
\begin{itemize}
\item We introduce elaTCSF with a spatial probability summation model \footnote{The code for the model and the datasets used to train it can be found on the project page: \url{https://www.cl.cam.ac.uk/research/rainbow/projects/elaTCSF/}.}, which accounts for eccentricity, luminance, and area, extending the industry flicker detection standard TCSF$_{IDMS}$. We also address past controversies regarding parafovea sensitivity peak.

\item We measure the visibility of flicker on a \ac{VRR} display. The measurements are combined with publicly available flicker detection data to calibrate and test elaTCSF. The dataset will be made publicly available. 

\item \edit{We demonstrate several applications of the model, including predicting safe refresh rate ranges for VRR displays, addressing VR headsets low-persistence flicker, and assisting in lighting design.} 
\end{itemize}

\section{Related Work}
\subsection{Temporal sensitivity and Critical Flicker Frequency} 
The neurons in the retina, thalamus, and subsequent stages of the visual pathway are sensitive to temporal variations in the retinal image \cite{Robson1966SpatialSystem,Breitmeyer1975TheResponse,Rucci2018TemporalSpace}. The sensitivity of human observers to temporal variations in light intensity has been extensively measured through psychophysical experiments~\cite{de1958research, kelly1961visual, Robson1966SpatialSystem, hartmann1979peripheral, tyler1987analysis, kong2018modelling}. Some researchers~\cite{porter1902contributions, hecht1933intermittent, hartmann1979peripheral, krajancich2021perceptual, chapiro2023critical} have focused on determining the frequency at which full-depth temporal modulation fuses to a steady light, known as \ac{CFF}. \ac{CFF} serves as a behavioral measure of temporal resolution \cite{Donner2021TemporalMeaning}. Understanding \ac{CFF} aids in predicting when the human visual system becomes insensitive to flicker, crucial for lighting systems design~\cite{watson1986temporal}. The Ferry–Porter law~\cite{porter1902contributions} explains the variation of CFF as a function of luminance and the Granit–Harper law~\cite{granit1930comparative} as the function of stimulus area.

\subsection{Flicker visibility metrics}
\citeN{watson20155} extended their prior flicker visibility metric  \cite{watson201164} to account for the effects of luminance.  They employed a bilinear TCSF model, fitted to the high-temporal-frequency limbs of the flicker detection measurements reported by \cite{de1958research}. Additionally, they assume that the Ferry-Porter law applies not only to CFF (contrast equal to one) but also to contrasts lower than one. This metric reports visibility in terms of Just Noticeable Differences (JNDs).

\citeN{Farrell1986AnFlicker} proposed a method to predict visible flicker in displays. It relies on the idea that, in a temporal amplitude sensitivity plot with double logarithmic axes, curves for different luminance levels converge to a common high-temporal-frequency asymptote (see \cite{kelly1961visual}, for a critical view, see \cite{Rider2019}), rendering CFF independent of adapting luminance. Note that this trend does not hold for contrast sensitivity curves.

\subsection{Variable Refresh Rate Flicker (VRR)}
A \ac{GPU} renders frames at varying rates, which do not align with a display's fixed refresh rate. Traditionally, a \ac{GPU} used a vertical synchronization signal (VSync) to ensure frames are displayed only when fully drawn. VSync, however, could cause skipped frames (stutter) and introduce unnecessary latency. To allow adaptive frame rate and reduce the latency, NVIDIA introduced \ac{VRR} technology in 2013. \ac{VRR} relies on the display holding pixel intensity for varying amounts of time, e.g., with capacitors in LCD panels \cite{slavenburg202046}. In \ac{VRR} mode, minor differences in display luminance at various refresh rates create low-frequency components in the Fourier domain, leading to visible flicker. Such flicker typically has low contrast and cannot be explained by the \ac{CFF} models and data. Our aim is to develop a model that can explain \ac{VRR} flicker.

\begin{figure}[t]
  \centering
      \includegraphics[width=\linewidth]{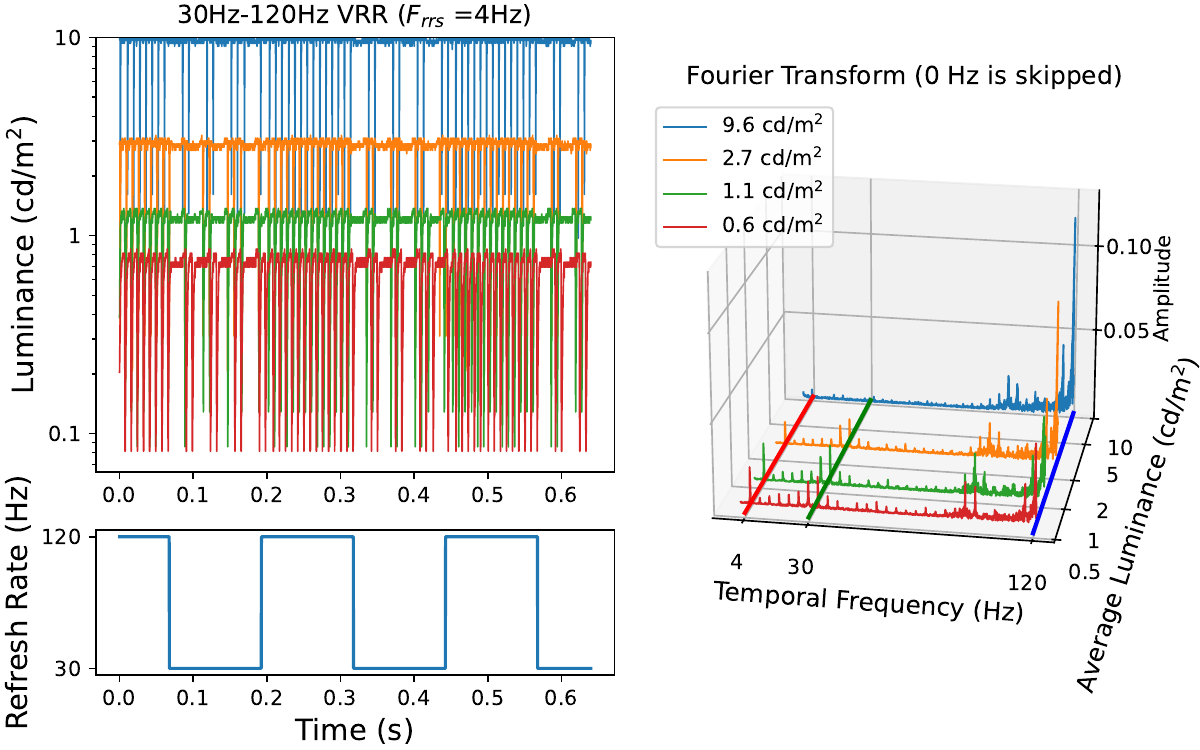}
  \caption{Left: Measurement of a VRR display, which alternates between 30\,Hz and 120\,Hz in short time intervals. The upper plot shows drops in luminance (caused by \ac{V-blank}) that vary in frequency depending on the refresh rate. The differences in the frequency of those drops cause small luminance differences and result in flicker. Right: The same signal in the frequency domain shows peaks at 4\,Hz (the frequency of the refresh rate change), 30\,Hz and 120\,Hz (caused by \acp{V-blank}).
  }
  \label{fig:vbi}
\end{figure}

\begin{figure}[t]
  \centering
  \includegraphics[width=\linewidth]{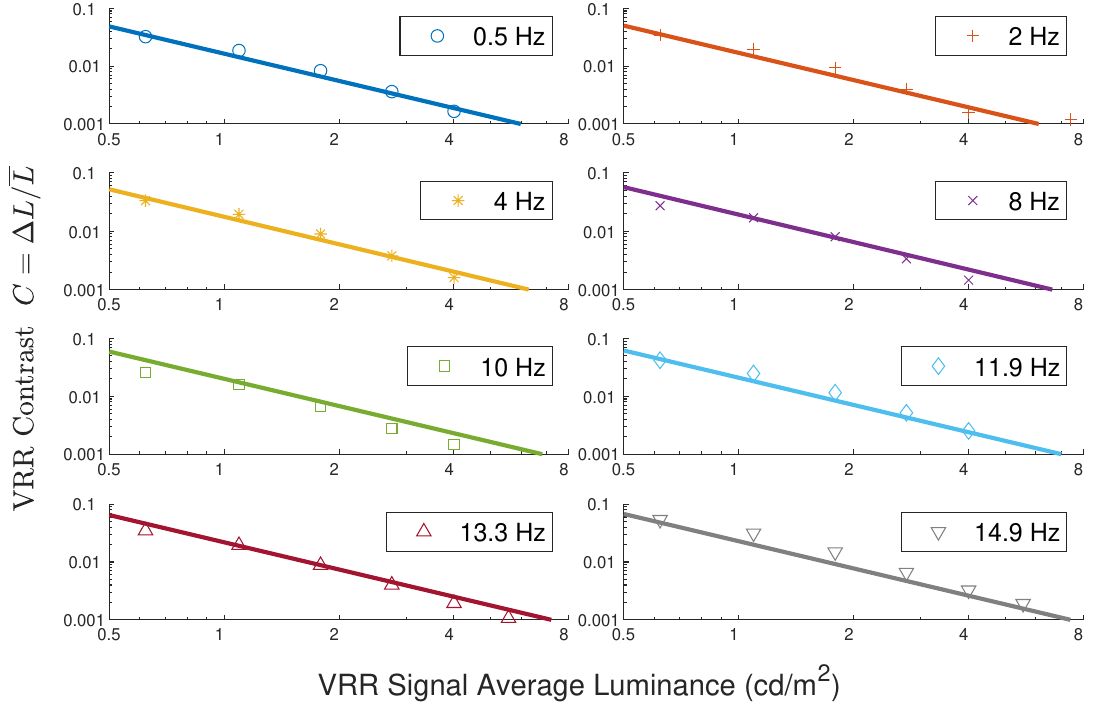}
  \caption{
The contrast of VRR flicker at varying stimulus luminance and temporal frequencies or refresh rate switch, $F_{\mathrm{rrs}}$ [Hz]. Luminance levels above 8\csdm are irrelevant to us because none of our VRR experimental data reaches above that luminance.}
  \label{fig:deltaL-L}
\end{figure}

\begin{figure}[ht]
  \centering
  \includegraphics[width=\linewidth]{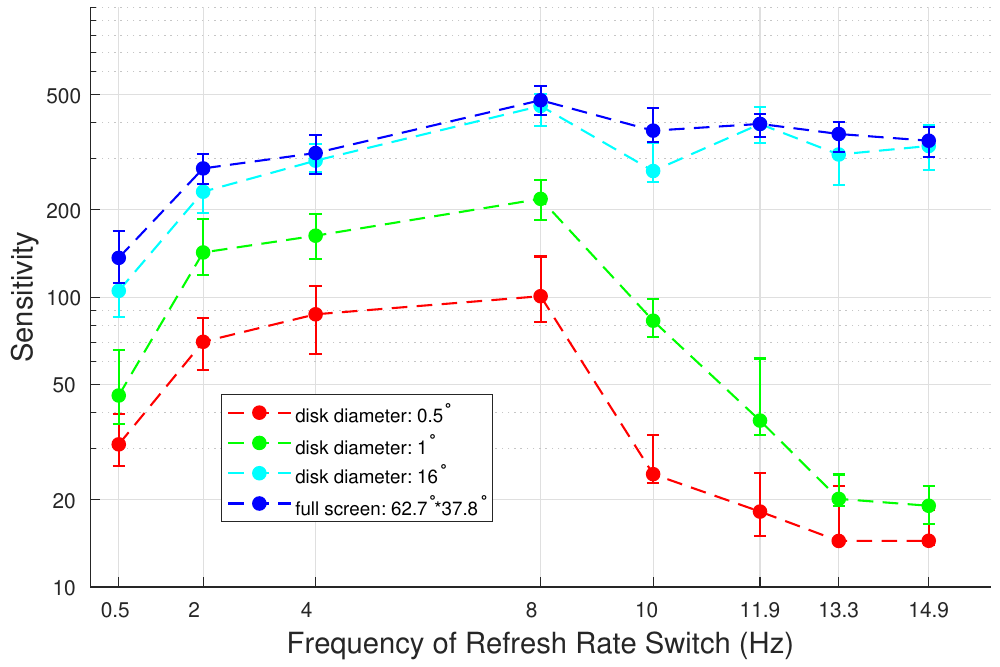}
  \caption{Our \ac{VRR} Flicker dataset, where each point represents the average sensitivity across all participants. The error bars indicate the upper and lower bounds derived from psychometric function fitting.}
  \label{fig:VRR_dataset}
\end{figure}

\begin{figure}[ht]
  \centering
  \includegraphics[width=\linewidth]{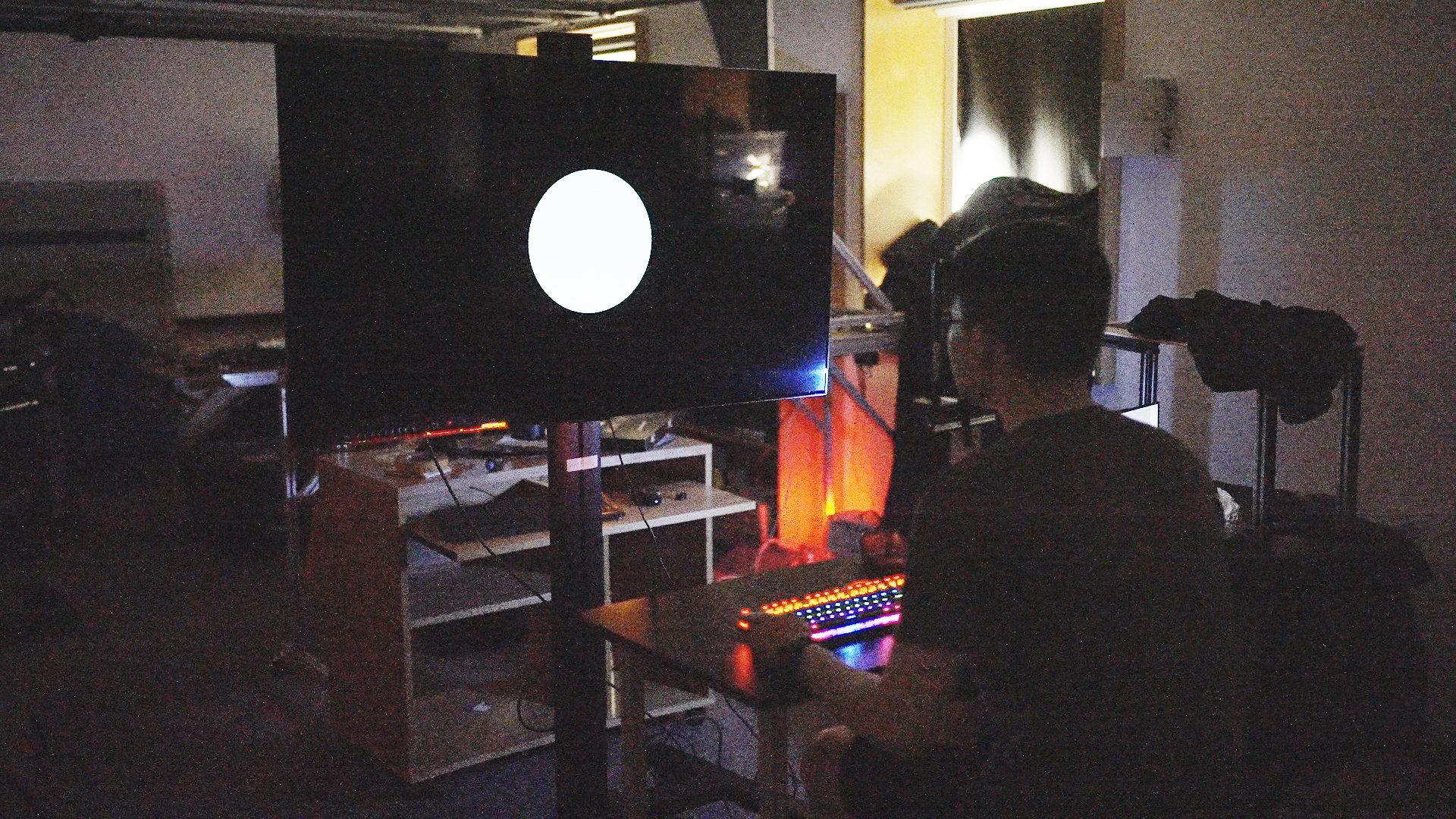}
  \caption{\edit{A photograph of the experimental setup. The experiment was taken in a dark room. The lights were added to take the photograph and were not present during the experiment.}}
  \label{fig:perceptual_visualization}
\end{figure}

\section{VRR Flicker Sensitivity Experiment}
\label{sec:vrr-flicker-experiment}
To model the visibility of flicker found in \ac{VRR} displays, we conducted measurements on a \ac{VRR}-capable (G-Sync) OLED display. Based on these measurements, we designed a psychophysical flicker detection experiment to measure human observers' sensitivity to \ac{VRR} flicker under varying conditions. 

\label{sec:VRR_Experimnet}
\paragraph{Display Equipment}
The measurements were conducted on an LG OLED G1 55" 4K Smart TV, chosen because it represents modern \ac{VRR} displays and exhibits \ac{VRR} flicker at low luminance levels. Moreover, its size is large enough to support flicker experiments over a very wide field of view. \edit{We measured the uniformity of the display and found the luminance difference between the central part and the edges to be less than 10\%.}

\paragraph{VRR Flicker Stimuli}
The stimuli were discs of varying sizes (0.5, 1, and 16 visual degrees in diameter) and a rectangular uniform field occupying the entire screen ($62.7^\circ \times 37.8^\circ$). The discs were shown on a black background. To induce flicker, the refresh rate was switched between 30 and 120\,Hz in a square-wave pattern (see \figref{vbi}-bottom-left). The frequency of refresh rate switch ($F_{\mathrm{rrs}}$) varied between 0.5 and 14.9\,Hz. 

\paragraph{Display calibration and measurements}
Since the primary cause of \ac{VRR} flicker is the subtle luminance differences across varying refresh rates, these differences result in low-frequency components in the Fourier domain, causing perceptible flicker. Therefore, flicker visibility is linked to the average screen luminance. We measured absolute display luminance levels with the Konica Minolta Chroma Meter CS-200. Temporal variation in luminance was measured with the Topcon Luminance Colorimeter RD-80SA, which provides a fast analog channel response of less than 80\,${\mu}s$. \figref{vbi}-left visualizes some temporal measurement results. The subtle luminance differences between 30\,Hz and 120\,Hz produce low-frequency components in the Fourier transform frequency domain, as shown in \figref{vbi}-right. We transform the signal into the frequency domain to identify the fundamental frequency (matching $F_{\mathrm{rrs}}$). The flicker contrast is calculated as the ratio of the modulation amplitude at the fundamental frequency to the mean luminance of the \ac{VRR} stimulus: $C = \frac{\Delta L}{\overline{L}}$.

Based on the above measurements, we can model flicker contrast as the function of display luminance and $F_{\mathrm{rrs}}$, as shown in~\figref{deltaL-L}. The flicker contrast did not depend on the size of the displayed pattern in our measurements.

\paragraph{Experimental Procedure}
\label{sec:subsec_vrr_procedure}
Due to the characteristics of \ac{VRR} flicker, we cannot directly control its contrast. Instead, we rely on the fact that the contrast of the flicker increases as we lower the luminance of the stimulus (see \figref{deltaL-L}). Therefore, the observers directly adjusted luminance and indirectly contrast in our experiments. 

The detection thresholds were measured in two stages. First, the observers were instructed to adjust the luminance of each \ac{VRR} stimulus until they could just detect flicker (method-of-adjustment). The adjusted luminance was then used as a starting point (a prior) for the 2-interval-forced-choice experiment. In this stage, the observers were presented with two intervals in random order: one containing a fixed refresh rate and one with a modulated refresh rate (\ac{VRR} flicker). The trials were controlled using the QUEST adaptive sampling method \cite{watson1983quest} (40 trails), implemented in PsychoPy~\cite{peirce2019psychopy2}. Responses were fitted to a psychometric function, and the contrast level at which a 0.75 correct response probability was reached was selected as the threshold contrast $C_t$ for a specific stimulus. The sensitivity $S$ is then computed as the reciprocal of $C_t$: $S = \nicefrac{1}{C_t}$. \edit{\figref{perceptual_visualization} shows a photograph of a participant using a chin rest when observing a stimulus on the display.}

\paragraph{Participants}
We recruited a total of 16 participants, divided into three groups. The first group consisted of 4 participants who completed experiments for all 8 $F_{\mathrm{rrs}}$ frequencies. The second group comprised 10 participants who only participated in experiments for the low-frequency range ($F_{\mathrm{rrs}}=$ 0.5, 2, 4, 8 Hz), while the third group consisted of 2 participants who solely took part in experiments for the high-frequency range ($F_{\mathrm{rrs}}=$ 10, 11.9, 13.3, 14.9 Hz). Through the t-test examining sensitivity across various temporal frequencies, we demonstrated that there was no significant deviation between the participants of the second and third groups ($t(3)=0.0066, p=0.9951, sd=0.2449$). \edit{The experiment was approved by the departmental ethics panel.}

\paragraph{Results}

The results are shown in \figref{VRR_dataset}. For signals of any size, sensitivity to flicker varies with temporal frequency, peaking around 8\,Hz. Stimuli with larger areas result in higher sensitivities. Additionally, for stimuli with small areas, sensitivity decreases rapidly as temporal frequency exceeds 8\,Hz, while such a decrease is much smaller for stimuli with larger areas. Note that due to the contrast-luminance correlation, the luminance associated with each data point varies with the reported sensitivity.

\begin{table*}[t]
\begin{center}
\caption{All flicker detection datasets utilized in our experiments. 
}
\label{tab:datasets}
\setlength{\tabcolsep}{2.0mm}{\scalebox{1}{

\begin{tabular}{c|c|c|c|c|c|c|c}
\toprule
Dataset        & \begin{tabular}[c]{@{}c@{}}Temporal Frequency \\ Hz\end{tabular} & \begin{tabular}[c]{@{}c@{}}Spatial Frequency \\ cpd\end{tabular} & \begin{tabular}[c]{@{}c@{}}Eccentricity \\deg \end{tabular} & \begin{tabular}[c]{@{}c@{}}Luminance \\ \csdm\end{tabular} & \begin{tabular}[c]{@{}c@{}}Area\\ deg$^2$\end{tabular} & Data & \begin{tabular}[c]{@{}c@{}} \edit{Data} \\ \edit{points}\end{tabular}  \\ \hline\hline
\citeN{hartmann1979peripheral}   & 7.77 - 61.18 & 0 & 0 - 60.37 & 0.7 - 70 & 0.2 - 7.07 & CFF & \edit{136}\\ \hline
\citeN{de1958research}A   & 23.97 - 64.23 & 0 & 0 & 0.16-1617.8 & 3.14 & CFF & \edit{7}\\ \hline
\citeN{krajancich2021perceptual}  & 23.01 - 94.41 & 0.01-0.57 & 0 - 55.04 & 3 - 190 & \textgreater 4.71 & CFF & \edit{36} \\ \hline
\citeN{chapiro2023critical}  & 31.88 - 51.21 & 0 & 0 - 20 & 10 - 8000 & 0.79 & CFF & \edit{30} \\ \hline
\edit{\citeN{de1958research}B}   & \edit{1.51 - 66.31} & \edit{0} & \edit{0} & \edit{0.16-1591} & \edit{3.14} & \edit{Sensitivity} & \edit{100}\\ \hline
\edit{\citeN{kelly1961visual}}  & \edit{1.6 - 75} & \edit{0.01} & \edit{0} & \edit{0.34 - 4928.7} & \edit{2827.4} & \edit{Sensitivity} & \edit{71}\\ \hline
\citeN{Snowden1995}  & 0.76 - 55.72 & 0.1 - 1 & 0 & 0.11 - 236 & 8.03 & Sensitivity & \edit{120}\\ \hline
\hline 
VRR Flicker (Ours)  & 0.5 - 14.9  & 0 & 0 & 0.47 - 4.13 & 0.2-2369.3  & Sensitivity &  \edit{32} \\ 
\bottomrule
\end{tabular}

}}
\end{center}
\end{table*}

\section{Model}
\label{sec:model}

To model flicker visibility, we extend Watson's \ac{TCSF}~\cite{watson201164} included in the \ac{IDMS}.  We decided not to extend existing \acp{CSF}, such as Barten's CSF \cite{barten1999contrast} or stelaCSF \cite{mantiuk2022stelacsf}, for three primary reasons. First, \acp{CSF} introduce spatial frequency dependency, posing challenges in integrating them into display flicker modeling~\cite{ashraf2023modelling}. Second, most sources of flicker tend to be of low spatial frequency (\tableref{datasets}). The sensitivity is mostly the function of stimulus size in the low-frequency range \cite{Savoy_McCann_1975}, and the dependence on the frequency only introduces unnecessary complexity. Furthermore, for high temporal frequencies that are relevant for flicker, sensitivity is mostly invariant to spatial frequency \cite{watson1986temporal} (spatial frequency only affects sensitivity at low temporal frequencies). 

\begin{figure}[t]
  \centering
  \includegraphics[width=\linewidth]{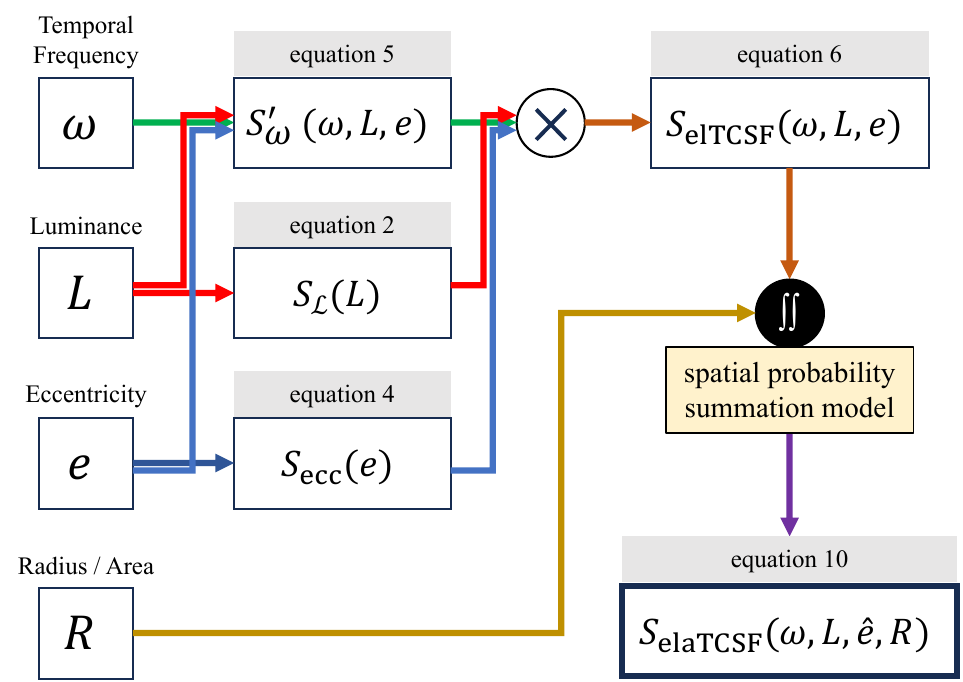}
  \caption{\edit{The inputs (left) and computation steps of elaTCSF.}}
  \label{fig:elaTCSF_diagram}
\end{figure}

The original TCSF$_{IDMS}$ only considers temporal frequency $\omega$:
\begin{equation}
S_{\omega, \mathrm{IDMS}}(\omega)=\left|\xi\left[(1+2 \mathrm{i} \pi \omega \tau)^{-n_{1}}-\zeta(1+2 \mathrm{i} \pi \omega \kappa \tau)^{-n_{2}}\right]\right|
\label{eq:TCSF_IDMS},
\end{equation}
where $\xi = 148.7$, $\tau = 0.00267$, $\kappa = 1.834$, $\zeta = 0.882$, $n_{1} = 15$, $n_{2} = 16$, which were fitted to the \citeN{de1958research} data. We extend it to account for three new dimensions: luminance, eccentricity, and area. The following sections explain how each new dimension is modeled. \edit{ \figref{elaTCSF_diagram} summarizes the computational steps.}

\subsection{Luminance} 
\label{sec:5.1}
In dim light, contrast sensitivity increases proportionally to the square root of retinal illuminance, in accordance with the DeVries-Rose law. Conversely, in bright light, sensitivity follows Weber’s law, remaining independent of illuminance~\cite{rovamo1995neural}. To incorporate those findings, we adopt the castleCSF~\cite{ashraf2024castlecsf} equation for the transient channel to model luminance sensitivity:
\begin{equation}
S_{\mathcal{L}}(L)=k_{1, \mathcal{L}}\left(1+\frac{k_{2, \mathcal{L}}}{L}\right)^{-k_{3, \mathcal{L}}}
\label{eq:luminance_1},
\end{equation}
where $k_{1...3, \mathcal{L}}$ are model parameters, which will be fitted.

The increase in luminance not only increases sensitivity but also shifts the peak of the TCSF towards higher temporal frequencies. Although not explicitly modeled, this phenomenon has been confirmed in the fitting results of~\cite{ashraf2024castlecsf} (Fig. 8 in castleCSF paper, (b,ii)). It suggests that human sensitivity to high temporal frequencies increases with higher luminance. This effect is modeled as a modification to the original TCSF$_{IDMS}$ function:
\begin{equation}
S_{\omega}(\omega, L) = S_{\mathcal{\omega}, \mathrm{IDMS}}\left(\frac{\omega}{b_{1, \omega}+k_{1, \omega}\mathrm{log}_{10}L }\right)
\label{eq:luminance_2},
\end{equation}
where $k_{1, \omega}$ and $b_{1, \omega}$ are the model parameter to be fitted.


\subsection{Eccentricity} 
\label{sec:5.2}
As mentioned in the Introduction, the change of sensitivity with eccentricity in the flicker detection task has been debated for over 120 years. Despite numerous attempts to explain this phenomenon, such as ~\citeN{rovamo1984critical}'s explanation by simultaneous scaling of stimulus area (M-scaling, retinal magnification) and illuminance (F-scaling), and ~\citeN{hartmann1979peripheral}'s proposition that larger area and higher luminance lead to the parafovea peak, there is still no universally accepted explanation.

Recent fMRI research on the human primary visual cortex suggests that the increase of flicker sensitivity in eccentricity is associated with the transient channel, which is sensitive to higher spatial frequencies. \citeN{horiguchi2009two} identified distinct spatial distributions of the sustained and transient channels. The transient channel exhibits maximal weighting in the parafovea. \citeN{himmelberg2019eccentricity} further confirmed that the peripheral visual field is more sensitive to higher frequency stimuli by analyzing the changes in a contrast semisaturation point C$_{50}$\footnote{C$_{50}$ is a measure of contrast sensitivity in the visual cortex, representing the contrast level at which 50$\%$ of the full response is achieved.} across different time frequencies in the foveal, parafoveal, and peripheral regions of the 
visual cortex. Based on these findings, we can infer that the sensitivity to high temporal frequencies should increase with eccentricity. At the same time, the contrast sensitivity data suggests that the peak of the \ac{CSF} decreases with eccentricity. These two trends can be reconciled only if the TCSF changes its shape with the eccentricity --- it becomes flatter (slope becomes lower) as we increase the eccentricity (see \figref{CSF_CFF}-(a)).



Eccentricity's influence on sensitivity can be divided into two aspects: the effect of eccentricity on sensitivity, which follows a simplified formulation of the pyramid of visibility \cite{watson2018field}:
\begin{equation}
S_{\mathrm{ecc}}(e)=10^{-k_{1,\mathrm{ecc}}e}
\label{eq:ecc_1},
\end{equation}
and the effect of the eccentricity on the slope $S_{\omega}$:
\begin{equation}
S_{\omega}^\prime(\omega, L,e) = S_{\mathcal{\omega}, \mathrm{IDMS}}\left(\frac{\frac{\omega-\omega_p}{1+k_{2, \omega}e}+\omega_p}{b_{1, \omega}+k_{1, \omega}\mathrm{log}_{10}L}\right)
\label{eq:ecc_2},
\end{equation}
where $k_{1,\mathrm{ecc}}$ and $k_{2, \omega}$ are the model parameters to be fitted, and $\omega_p = -2$\,Hz is a factor to control the temporal frequency peak shift. Then, we can construct the base function elTCSF for  the subsequent spatial probability summation model:
\begin{equation}
S_{\mathrm{elTCSF}}(\omega ,L,e) = S_{\mathcal{L} }(L)S_{\mathrm{ecc}}(e)S_{\omega}^\prime(\omega ,L,e)\,.
\label{eq:elTCSF}
\end{equation}


\subsection{Area and Spatial Probability Summation Model}
\label{sec:5.3}
Stimulus area (size) significantly impacts sensitivity, as larger stimuli activate more retinal cells. However, existing methods \cite{barten1999contrast,mantiuk2022stelacsf} often treat area as a separate parameter, assuming that eccentricity and area effects on stimuli are independent. This assumption becomes unreasonable when dealing with very large stimuli, such as a full-screen flicker in our \ac{VRR} dataset, which covers a significant range of eccentricities. 
To account for varying eccentricity across the visual field, we use a spatial probability summation model. Specifically, we treat the visual field as a continuous function of contrast and integrate the product of contrast and sensitivity over the window $w{\times}h$ (in degrees):
\begin{equation}
E=\int_{-\frac{w}{2}}^{\frac{w}{2}} \int_{-\frac{h}{2}}^{\frac{h}{2}} (c(x, y) S(x, y))^{\beta} dx\,dy
\label{eq:area_1},
\end{equation}
where $\beta$ is the probability summation exponent (will be fitted to the data), $c(x, y)$ represents the stimulus contrast, which is independent of spatial position ($c(x,y) \equiv c$) in all datasets used. Because all datasets used have sufficiently long durations, we disregard the impact of duration. For circular stimuli (disks), we can employ polar coordinates to simplify the spatial probability summation model:
\begin{equation}
d(\hat{e}, r, \theta) = \sqrt{\hat{e}^{2}+r^{2}+2\,\hat{e}\,r\,\cos (\theta)}
\label{eq:area_2},
\end{equation}
\begin{equation}
E(\omega, L, \hat{e}, R)=c^{\beta} \int_{0}^{2\pi} \int_{0}^{R} S_{\mathrm{elTCSF}}^{\beta}\left(\omega, L, d(\hat{e}, r, \theta)\right) r\, dr \, d\theta
\label{eq:area_3},
\end{equation}
where $\hat{e}$ is the distance between the disk center and the fixation point (in visual degrees), and $R$ is the radius (in visual degrees). When the summation reaches the threshold $E_{thr}$, the flicker can be detected, thus the complete $S_{\mathrm{elaTCSF}} = 1 / c$ can be expressed as:
\begin{equation}
S_{\mathrm{elaTCSF}}(\omega, L, \hat{e}, R) = \left(\frac{\int_{0}^{2\pi} \int_{0}^{R} S_{\mathrm{elTCSF}}^{\beta}\left(\omega, L, d(\hat{e}, r, \theta)\right) r\, dr \, d\theta}{E_{t h r}}\right)^{\frac{1}{\beta}}
\label{eq:elaTCSF}
\end{equation}
where $E_{thr}$ is the parameter that is fitted separately per each dataset.

\paragraph{CFF prediction} \ac{CFF} is the temporal frequency at which the sensitivity is 1. The model from \eqref{elaTCSF} cannot be analytically inverted and solved for $\omega$. Therefore, to find the \ac{CFF}, we use numerical root finding in the range between 8 and 200\,Hz. 

\section{Model Fitting and Comparison}
\label{sec:model_Fit_Compare}
In this section, we present the optimization process for all parameters of elaTCSF and compare its accuracy with the existing models.

\subsection{Model Fitting}

\label{sec:subsec_model_fitting}
We curated numerous existing flicker detection datasets, shown in Table~\ref{tab:datasets}. We excluded the data points with spatial frequencies exceeding 1\,cpd (as elaTCSF is meant for low spatial frequencies) and those with low luminance ($L$ < 0.1\csdm). In total, \edit{8} datasets were used to train the parameters of our elaTCSF. Our optimization loss function is:
\begin{equation}
\mathcal{L}=\frac{1}{N} \sum_{d} \sum_{i}\left(\log _{10} S_{i, d}-s_{d} \log _{10} \tilde{S}_{i, d}\right)^{2}+\frac{\lambda}{D} \sum_{d}\left(\log _{10} s_{d}\right)^{2}
\label{eq:loss_1},
\end{equation}
where $d = 1...D$ is the dataset index, $i$ is the stimulus index in a dataset, $N$ is the total number of stimuli, $S_{i, d}$ and $\tilde{S}_{i, d}$ are the reference and predicted sensitivity values. 
$s_d$ is a per-dataset scaling factor that accounts for the difference between the datasets (differences in protocols, group of observers, etc.), as explained in stelaCSF paper\cite[sec.~6]{mantiuk2022stelacsf}. \edit{In summary, the first term of the loss function is responsible for data fitting, while the second term ensures that the per-dataset scaling factor $s_d$ remains close to 1.}
In all our experiments, we fix $s_{d} = 1$ for our VRR dataset. We set $\lambda$ to 0.001. The quasi-Newton method implemented in Matlab’s \textit{fminunc} function is used for optimization. The parameters of the fitted elaTCSF are listed in Table~\ref{tab:parameters}.


\begin{table}[t]
\begin{center}
\caption{The fitted parameters of elaTCSF.}
\label{tab:parameters}
\setlength{\tabcolsep}{1.4mm}{\scalebox{1}{
\begin{tabular}{c|c}
\toprule
Part & Parameters \\ \hline\hline
\edit{TCSF$_{IDMS}$} & \begin{tabular}[c]{@{}c@{}} \edit{$\xi = 154.133$, $\tau = 0.00292069$, $\kappa = 2.12547$}, \\ \edit{$\zeta = 0.721095$, $n_{1} = 15$, $n_{2} = 16$}\end{tabular} \\ \hline
Luminance & \begin{tabular}[c]{@{}c@{}} $k_{1, \mathcal{L}}$ = \edit{1.76801}, $k_{2, \mathcal{L}}$ = \edit{1.62402},  $k_{3, \mathcal{L}}$ = \edit{0.533781},  \\ $k_{1, \omega}$ = \edit{0.222269}, $b_{1, \omega}$ = \edit{0.6678}\end{tabular} \\ \hline
Eccentricity & $k_{1,\mathrm{ecc}}$ = \edit{0.0330933}, $k_{2, \omega}$ = \edit{0.0341811} \\ \hline
Area &$E_{thr}$ = \edit{6.52801}, $\beta$= \edit{3.80022} \\ \bottomrule
\end{tabular}

}}
\end{center}
\end{table}

\subsection{Model Comparison}
\label{sec:subsec_model_comparison}
\begin{table}[t]
\begin{center}
\caption{Comparisons. S-RMSE is reported only for datasets using sensitivity as the evaluation metric, while CFF-RMSE is reported only for datasets using CFF as the evaluation metric. Regular font numbers correspond to the results of the entire dataset, while small font magenta-colored numbers represent the results of 5-fold cross-validation, indicating the mean and standard deviation across all folds. Barten's CSF (HTF) and stelaCSF (HTF) are the updated versions from \cite{bozorgian2024spatiotemporal}.
}
\label{tab:comparisons}
\setlength{\tabcolsep}{1.4mm}{\scalebox{1}{

\begin{tabular}{c|c|c}
\toprule
CSF   Model & S-RMSE $\mathcal{E}_{s}$ {[}dB{]} & CFF-RMSE $\mathcal{E}_{\omega}$ \\ \hline\hline
TCSF$_{IDMS}$ & \edit{10.33 \validation{10.38 ± 1.64}} & \edit{13.52 \validation{13.93 ± 1.14}} \\ \hline
Barten's CSF~\cite{barten1999contrast} & \edit{5.79 \validation{6.08 ± 0.68}} & \edit{15.36 \validation{16.03 ± 7.77}} \\ \hline
stelaCSF~\cite{mantiuk2022stelacsf} & \edit{6.13 \validation{6.30 ± 0.36}} & \edit{11.47 \validation{11.82 ± 2.09}} \\ \hline
Barten's CSF (HTF) & \edit{4.58 \validation{4.82 ± 0.51}} & \edit{9.06 \validation{10.09 ± 1.41}} \\ \hline
stelaCSF (HTF) & \edit{6.05 \validation{6.30 ± 0.34}} & \edit{11.75 \validation{13.77 ± 2.25}} \\ \hline
castleCSF~\cite{ashraf2024castlecsf} & \edit{5.45 \validation{5.58 ± 0.55}} & \edit{13.07 \validation{14.70 ± 4.53}} \\ \hline\hline
elaTCSF (ours) & \edit{\textbf{3.50 \validation{3.73 ± 0.67}}} & \multicolumn{1}{l}{\edit{\textbf{8.95 \validation{9.07 ± 1.55}}}} \\ \bottomrule
\end{tabular}

}}
\end{center}
\end{table}


Following the stelaCSF~\cite{mantiuk2022stelacsf} validation protocol, we perform five-fold cross-validation within each dataset, utilizing all datasets for both training and testing. We also follow the approach of other works~\cite{ahumada2018dual, watson2005standard}, reporting results for the entire dataset without a training/testing split. We report two error measures: \ac{RMSE} of contrast sensitivity in dB units $\mathcal{E}_{s}$ (S-RMSE, see \cite[Eq.~19]{mantiuk2022stelacsf}) and the RMSE of \ac{CFF}  $\mathcal{E}_{\omega}$ (CFF-RMSE) in Hz. 
\edit{In \tableref{comparisons}} we report S-RMSE results \edit{for flicker detection datasets (labelled as ``Sensitivity'' in \tableref{datasets})}~\footnote{Notably, as elaTCSF does not consider spatial frequency, we used \cite{Snowden1995} dataset exclusively as the training set and excluded it during model comparison.} and CFF-RMSE results on \ac{CFF} datasets \edit{(labelled as ``CFF'' in \tableref{datasets})}.

Overall, elaTCSF outperforms the existing advanced CSF models for both error metrics. 
\edit{Barten’s CSF (HTF) shows the best performance among the current models for the S-RMSE and CFF-RMSE.} We have visualized the resulting fits in Figures~\ref{fig:deLange_CFF}--\ref{fig:ours_vrr_1}. 

\edit{The fitting results of elaTCSF on the \citeN{de1958research} (\figref{deLange_S}), \citeN{kelly1961visual} (\figref{kelly}), and our VRR datasets (\figref{ours_vrr_1}) demonstrate its accuracy in predicting sensitivity values across temporal frequencies under varying luminance and stimulus size conditions. \figref{deLange_S} and \figref{kelly} illustrate that our elaTCSF model is applicable across a wide range of luminance levels. The results on our dataset (\figref{ours_vrr_1}), in particular, validate the effectiveness of our spatial probability summation model, as elaTCSF performs well in predicting sensitivity for both large and small stimuli.}

The plots showing the fits to the datasets of \citeN{hartmann1979peripheral} (\figref{hartmann}) and \citeN{krajancich2021perceptual} (\figref{krajancich}) demonstrate that elaTCSF correctly predicts the increase of \ac{CFF} in the peripheral visual field. Both datasets reveal CFF peaks within the periphery (eccentricities of 10 to 30 degrees). Although Barten's CSF (HTF)~\cite{bozorgian2024spatiotemporal} attempts to model these peaks, it predicts the peak eccentricities to be much lower than those observed experimentally. In contrast, our elaTCSF accurately models the parafoveal CFF peaks. While there are deviations at very low luminance and high eccentricities (\figref{hartmann}), the predictions made by elaTCSF are sufficiently accurate for practical applications.

 The datasets of \citeN{de1958research} (\figref{deLange_CFF}) and \citeN{chapiro2023critical} (\figref{hvei_cff}) demonstrate the relationship between \ac{CFF} and luminance, including very high luminance levels (up to 8000\csdm), which is beneficial for HDR display design. While elaTCSF outperforms existing models, it is inconsistent with the dataset of \citeN{chapiro2023critical}. According to their data, the CFF values at fovea (0 degree) are higher than those in the parafoveal regions (10, 20 degrees), which is inconsistent with the trends observed in the datasets of \citeN{krajancich2021perceptual} and \citeN{hartmann1979peripheral}. The factors that could cause this inconsistency between the datasets are unknown and, consequently, cannot be modeled.



\section{Applications}
\edit{The primary application elaTCSF addresses is VRR flicker, which we measured and then validated our model on. Below in \secref{vrr-app} we demonstrate how VRR flicker can be mitigated in practice. We also show how elaTSCF can be used to predict flicker in VR/AR headsets (\secref{low-pers-flicker}), and how it can serve as a better flicker model for lighting design (\secref{app-lighting-design}). The latter two application, however, are not valiated. 
}

\subsection{Prediction of Frame Rate Range for VRR Displays}
\label{sec:vrr-app}
\setlength{\belowcaptionskip}{-10pt}

\begin{figure}[t]
  \centering
  \includegraphics[width=\linewidth]{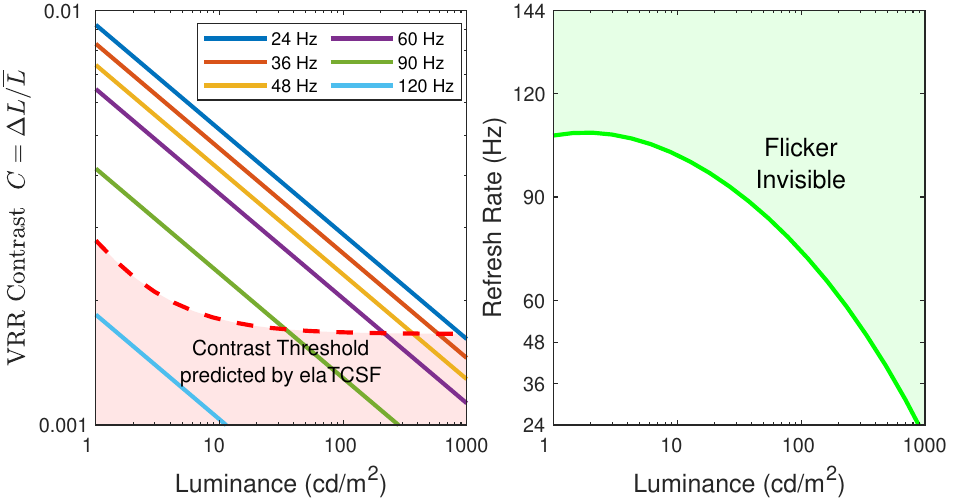}
  \caption{Left: Measured/simulated contrast of VRR flicker when switching between 144\,Hz and given refresh rate (lines). The dashed line is the detection threshold predicted by elaTCSF for the given display. Right: The pale green area is the range of refresh rates in which the display can operate without visible flicker. The lower bound of that range is the intersection of the contrast lines with the threshold, in the plot on the left. Note that higher luminance makes VRR flicker less visible because of the reduced VRR-flicker contrast.}
  \label{fig:application_vrr}
\end{figure}

\begin{figure}[t]
  \centering
  \includegraphics[width=\linewidth]{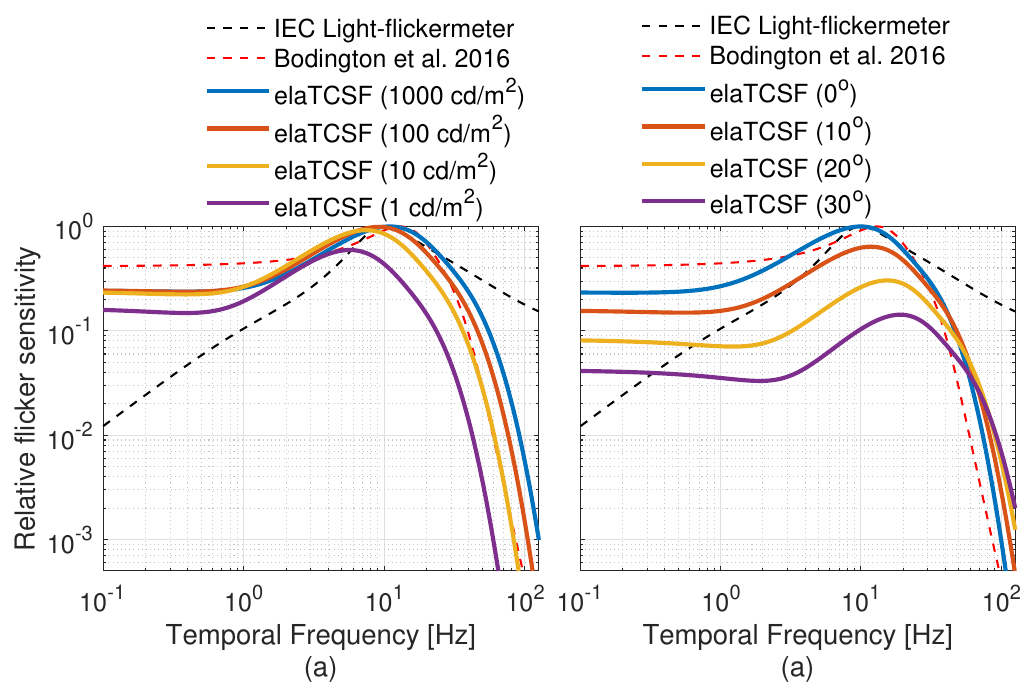}
  \caption{Comparison of relative flicker sensitivity functions from the IEC Light-flickermeter and \cite{bodington2016flicker} with the predictions of elaTCSF model at (a) various luminance levels for a foveal visual field of 1000\,degrees$^2$, and (b) fovea and different peripheral eccentricities for visual fields of 100\,degrees$^2$ and 1000\csdm. The model predictions are normalised with respect to the maximum sensitivity from  1000\csdm condition for plot (a), and the foveal condition for plot (b). }
  \label{fig:application_lighting}
\end{figure}

\setlength{\belowcaptionskip}{0pt}

A common method to mitigate \ac{VRR} flicker is to limit the operational range of frame rates when \ac{VRR} is enabled. For instance, even if the monitor supports the range from 24\,Hz to 144\,Hz, it is restricted to 40\,Hz to 144\,Hz in the \ac{VRR} mode using low frame rate compensation (LFC). This approach reduces flicker visibility by minimizing potential luminance differences between frame rates. Currently, without a robust model, engineers manually adjust the refresh rate range until the flicker cannot be noticed.

Our elaTCSF can calculate the flicker-free refresh rate range for \ac{VRR} displays at different luminance levels.
In this example, we will simulate a 27-inch 24--144\,Hz display. First, following our procedure from \secref{vrr-flicker-experiment}, we need to measure for each luminance level the contrast introduced by switching from the 144\,Hz (maximum) to any other frame rate. Such contrast for our simulated display is shown as lines in \figref{application_vrr}. Then, we use elaTCSF to find the detection threshold for the display across the luminance range, shown as a dashed line in \figref{application_vrr}-left. We use the maximum sensitivity across all temporal frequencies to ensure conservative thresholds. The intersection of the VRR-flicker contrast lines with the threshold gives us the lower bound of the refresh rate, shown as the green line in \figref{application_vrr}-right. The pale green area indicates the range of refresh rates in which \ac{VRR} can operate without introducing visible flicker. 


\subsection{Low persistence flicker in VR headsets}
\label{sec:low-pers-flicker}
VR headsets display world-locked content that can move rapidly with head motion. Because of that, they are prone to hold-type blur \cite{rao2024display}. Such blur can be much reduced using a low-persistence display, which keeps the image on for a fraction of the frame duration (e.g., 2\,ms for a 7\,ms frame) and the display remains black for the remaining time. This introduces the stroboscopic effect, making the image appear sharper, but it can introduce visible and uncomfortable high-contrast flicker. Currently, VR manufacturers rely solely on empirical evidence to set the lower limit for refresh rates, such as Meta's assertion~\cite{rao2024display} that 90\,Hz is the minimum refresh rate for a comfortable user experience.


As our elaTCSF accounts for all relevant factors, we can predict the minimum refresh rate of a VR headset that ensures the low-persistence flicker remains invisible. Based on publicly available data regarding device's FoV, we predicted CFF for three headsets: Apple Vision Pro, Microsoft HoloLens 2 and Meta Quest 3. We plot CFF as the function of luminance in the \figref{teaser}-right. The predicted values assume that the display is showing a uniform field of a given luminance. If the display's refresh rate is above the CFF line, the low-persistence flicker is unlikely to be visible. The plots show how much the refresh rate would need to increase if we wanted to increase the display peak luminance without introducing flicker. 





\subsection{Application in lighting design}
\label{sec:app-lighting-design}

Flicker in lighting systems has long been a concern, initially identified with old incandescent bulbs. While advances in lighting technology, particularly with LEDs, have mitigated some flicker issues, they have not been completely resolved. Modern LED lights use a range of driver circuits to manage voltage fluctuations and flicker can still be present due to the type of LED driver used and its handling of voltage fluctuations \cite{collin2019light}. Changes in voltage supply can still produce perceptible flicker, especially with less efficient driver circuits. LED products can also exhibit flicker at dim light levels or during transitions between dimming states \cite{poplawski2013flicker}. The current standards for estimating the flicker index of light sources, such as those set by the International Electrotechnical Commission (IEC), rely on human contrast sensitivity measurements developed for incandescent lighting. These standards do not fully account for the differences in flicker sensitivity under varying lighting conditions \cite{compatibility2010part, international2020equipment}. Our elaTCSF model, supported by recent measurements can be used to update the current standards. A recent work by \cite{kukavcka2023comparison} highlighted the importance of incorporating human TCSF measurements in the light flicker index (LFI) for different light sources.

\figref{application_lighting} shows the prediction of our model versus the filter used in the IEC standard and from another TCSF measurement \cite{bodington2016flicker}. The filter used in IEC standard is from \cite{drapela2010light} and the measurements from \cite{bodington2016flicker} have been used in a recommendation to update the current lighting standards \cite{recommends2015recommended}. Both TCSF measurements from the literature do not show any variation with luminance, eccentricity, etc., whereas our model is able to predict changes in flicker sensitivity with different viewing conditions. This capability can be used to update lighting flicker index measurements for different conditions, providing 
a more comprehensive and perceptually-accurate framework for assessing flicker in modern lighting systems. 

\section{Conclusions}

A flickering light source, such as a display or lighting, can be very annoying, cause eye strain and headaches, and is a main concern in display and lighting design. A certain amount of flicker is unavoidable. Therefore, the main question is what is the largest contrast or the smallest temporal frequency at which the flicker becomes invisible. To that end, we proposed \ourmethod{}, which can predict both. It accounts for all main factors that influence flicker perception: temporal frequency, luminance, eccentricity and size. This is a significant improvement over existing models, which cannot predict the threshold contrast (e.g., \acp{CFF}), do not account for all relevant factors (e.g., \ac{IDMS} TCSF), or do not offer sufficient accuracy (e.g., stelaCSF). \ourmethod{} is fitted to and tested against \edit{8} different datasets with both \ac{CFF} and sensitivity measurements, one of which we collected specifically to address the prominent issue of flicker in \ac{VRR} displays. \ourmethod{} is built on established psychophysical models, such as Watson's TCSF, or the spatial probability summation. This choice was made to avoid overfitting given the sparsity of available psychophysical data. Even if a better fit can be found with a polynomial function or a neural network, such a function is unlikely to generalize to the conditions outside the training dataset.

The main limitation of \ourmethod{} is that it can only predict flicker for low-spatial-frequency or large patterns, as it does not account for spatial frequency. We found that this compromise was necessary to achieve both good quantitative and qualitative predictions, and because the previous attempts to extend more complex \acp{CSF} were not fully successful \cite{bozorgian2024spatiotemporal}. \edit{elaTCSF is trained and validated on VRR flicker data as this is the focus of our work. The model is yet to be validated on other applications, including flicker of low-persistence displays and flicker in lighting design.}  

\begin{acks}
We would like to thank anonymous reviewers for their feedback and suggestions, and experiment participants for their contribution to this project. 
\end{acks}



\bibliographystyle{ACM-Reference-Format}
\bibliography{elaTCSF,refs}

\appendix

\clearpage

\begin{figure}[H]
  \centering
  \includegraphics[width=\linewidth]{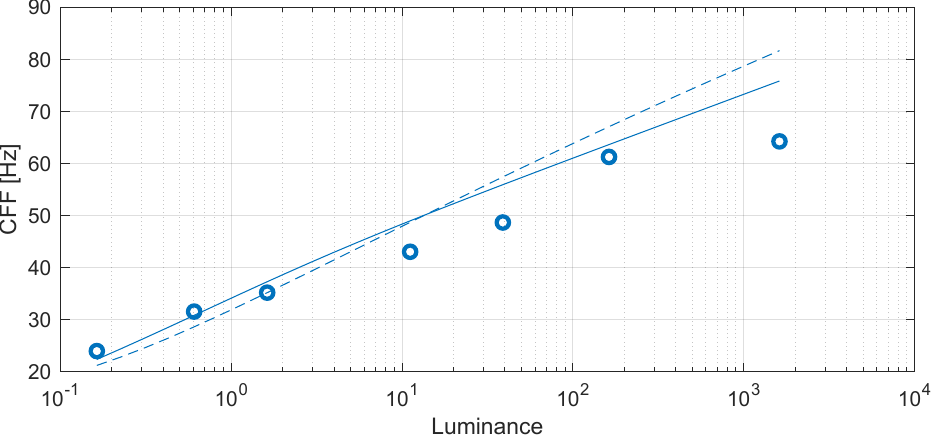}
  \caption{Predictions for \cite{de1958research} dataset (CFF). Continuous lines: elaTCSF; dashed lines: Barten's CSF (HTF).\vspace{-8pt}}
  \label{fig:deLange_CFF}
\end{figure}

\begin{figure}[H]
  \centering
  \includegraphics[width=\linewidth]{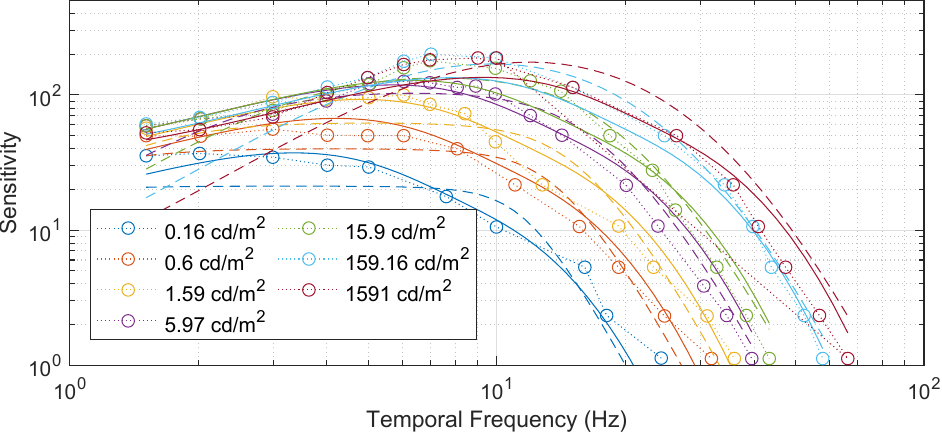}
  \caption{\edit{Predictions for \cite{de1958research} dataset (Sensitivity). Continuous lines: elaTCSF; dashed lines: Barten's CSF (HTF).\vspace{-8pt}}}
  \label{fig:deLange_S}
\end{figure}

\begin{figure}[H]
  \centering
  \includegraphics[width=\linewidth]{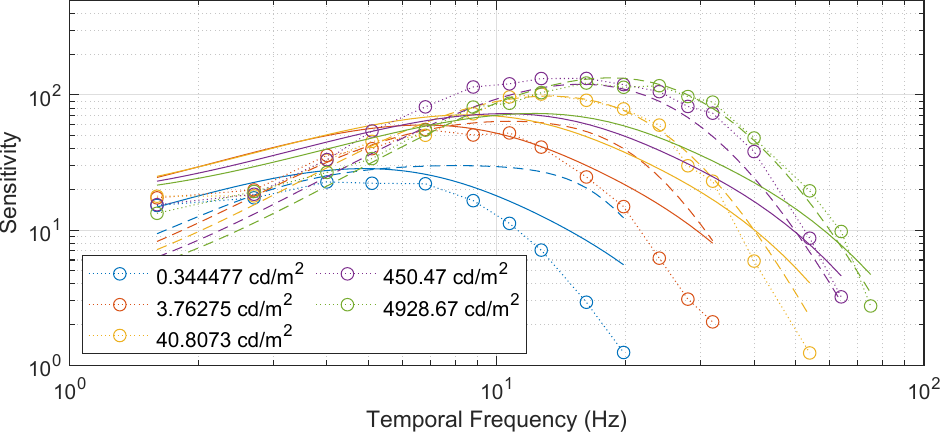}
  \caption{\edit{Predictions for \cite{kelly1961visual} dataset. Continuous lines: elaTCSF; dashed lines: Barten's CSF (HTF).\vspace{-8pt}}}
  \label{fig:kelly}
\end{figure}

\begin{figure}[H]
  \centering
  \includegraphics[width=\linewidth]{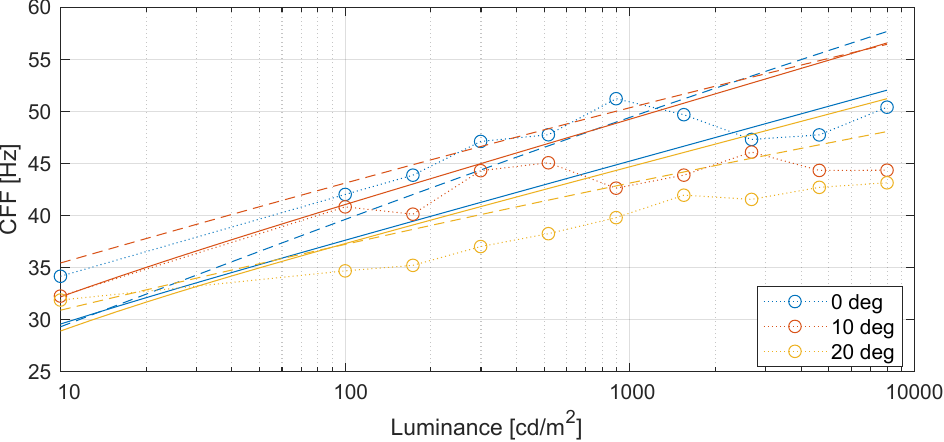}
  \caption{Predictions for \cite{chapiro2023critical} dataset. Continuous lines: elaTCSF; dashed lines: Barten's CSF (HTF).}
  \label{fig:hvei_cff}
\end{figure}

\begin{figure}[H]
  \centering
  \includegraphics[width=0.9\linewidth]{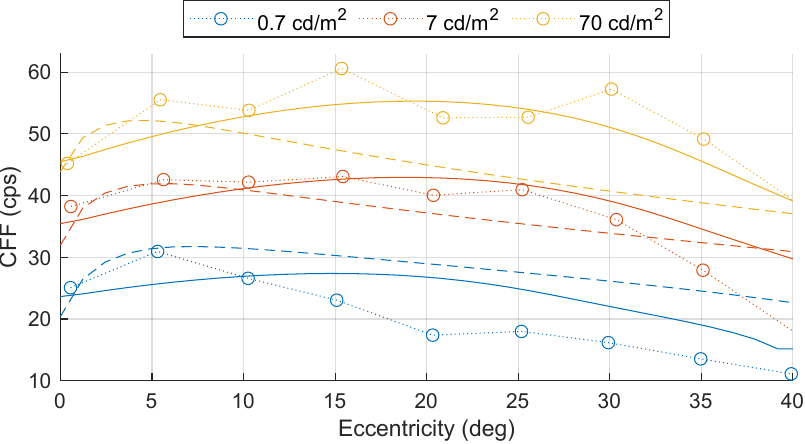}
  \caption{Predictions for \cite{hartmann1979peripheral} dataset. Continuous lines: elaTCSF; dashed lines: Barten's CSF (HTF).}
  \label{fig:hartmann}
\end{figure}

\begin{figure}[H]
  \centering
  \includegraphics[width=0.9\linewidth]{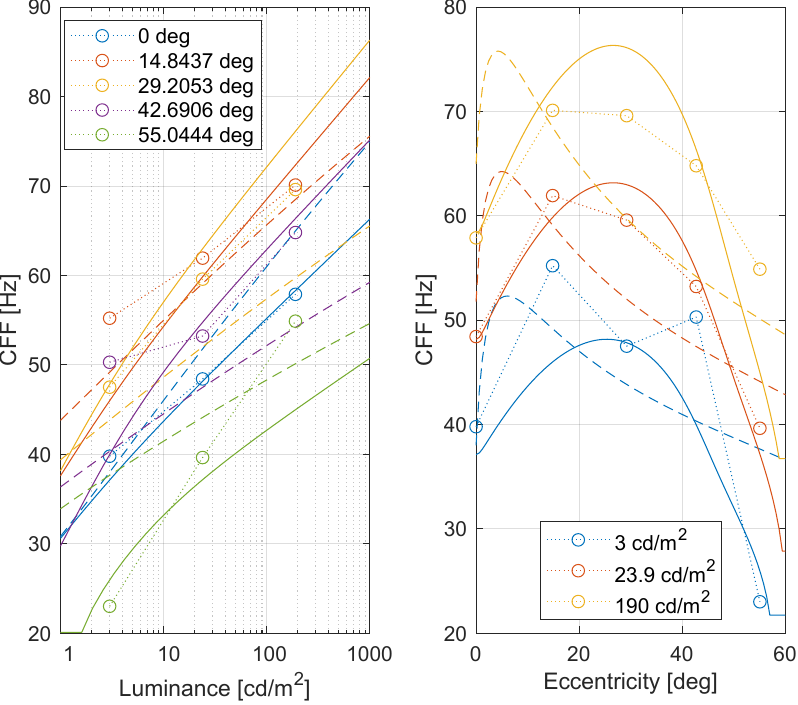}
  \caption{Predictions for \cite{krajancich2021perceptual} dataset. Continuous lines: elaTCSF; dashed lines: Barten's CSF (HTF).}
  \label{fig:krajancich}
\end{figure}

\begin{figure}[H]
  \centering
  \includegraphics[width=\linewidth]{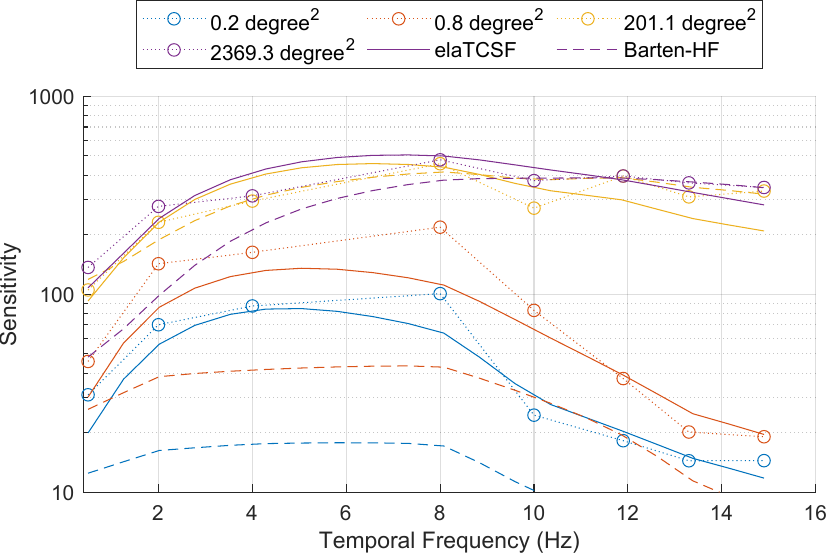}
  \caption{Predictions for our VRR dataset (Temporal Frequency). \edit{Continuous lines: elaTCSF; dashed lines: Barten's CSF (HTF)~\cite{bozorgian2024spatiotemporal}.}}
  \label{fig:ours_vrr_1}
\end{figure}

\begin{figure*}[ht]
  \centering
  \includegraphics[width=\linewidth]{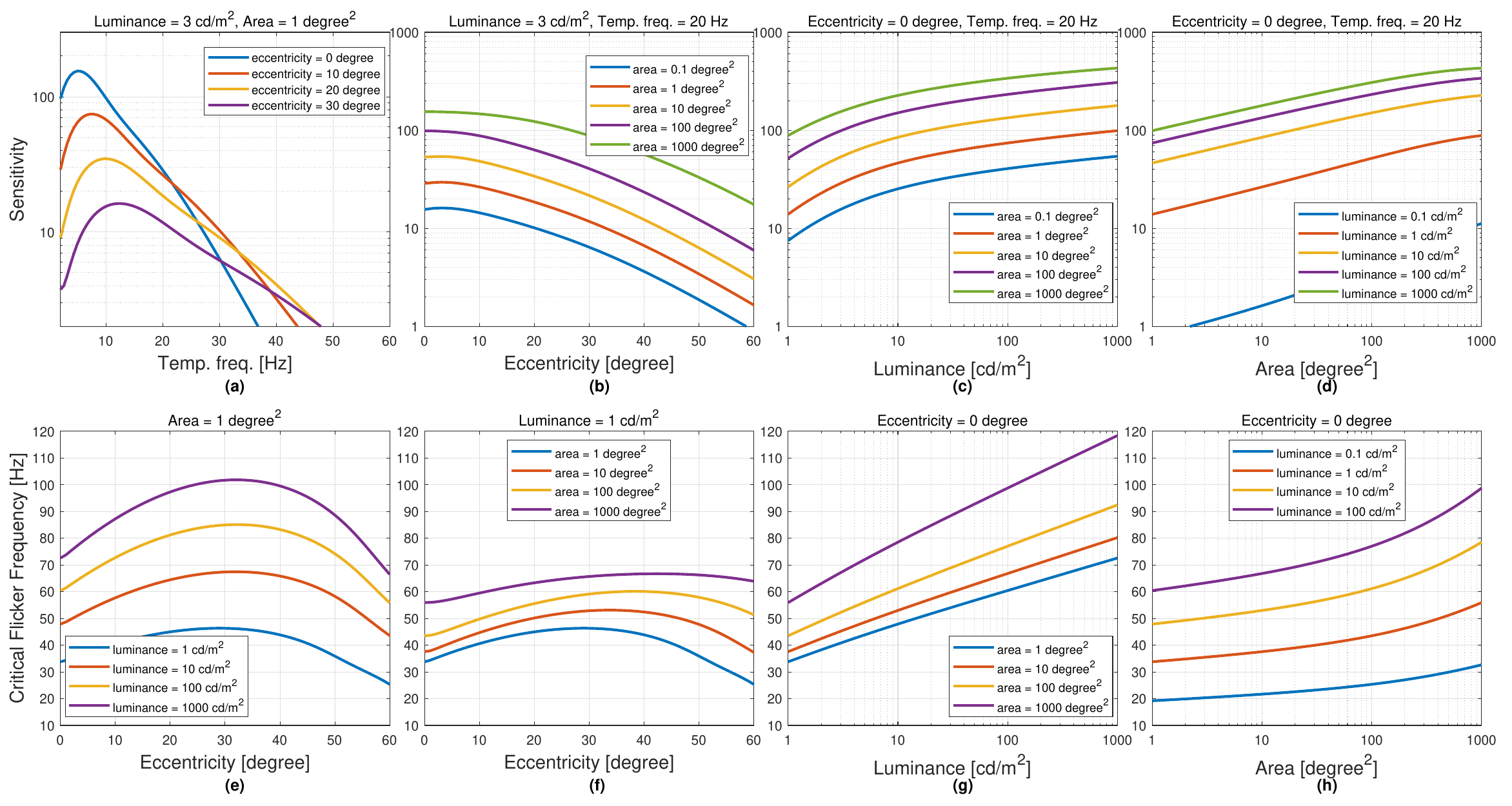}
  \caption{The sensitivity (first row) and critical flicker frequency (second row) computed for elaTCSF under different conditions.}
  \label{fig:CSF_CFF}
\end{figure*}


\end{document}